# Comment on 'Semiconductor nanocrystals: structure, properties, and band gap engineering' [Acc. Chem. Res. Vol 43 (2) pp 190, (2010)]

Sesha Vempati*

UNAM-National Nanotechnology Research Center, Bilkent University, Ankara, 06800, Turkey.

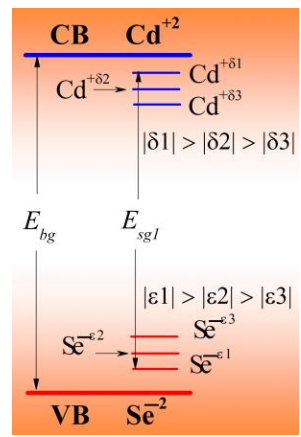

**ABSTRACT:** Surface traps and associated emission in quantum dots (QDs) sought a lot of research attention because of the fundamental interests apart from their influence on the emission characteristics. In ref [Acc. Chem. Res. Vol 43 (2) pp 190, 2010] the hole traps ($h$-traps) are depicted close to the conduction band (CB) for CdSe QDs while discussing the emission mechanism. However, notably electron traps ($e$-traps) are close to CB and $h$-traps are supposedly close to the valance band, especially in anion rich CdSe QDs. Although such emission (so called deep-trap) is a well known phenomenon, the energetic locations of these traps and the associated discrepancy require further attention. Hence the distinction between $e$, $h$-traps and their energetic location within the band gap is addressed in a general context, which is essentially a revisit to the 'surface states'. Finally this general description is put to the context of surface states of CdSe QDs.

Keywords: surface defects, quantum dots, cadmium selenide, hole traps, electron traps

## I. DISCUSSION

We start the discussion by assuming an ionic crystal of the form $M_nX_m$. For the sake of simplicity, broadening of the conduction band (CB), valance band (VB) and the presence of the chemisorbed ions are not considered, which of course does not undermine the argument in anyway. Also the passivation of surface states is not considered as we are interested in discussing their presence and energetic location. In the bulk crystal, the anions and cations form the VB ($E_{bX}$) and CB ($E_{bM}$) levels, respectively (Figure 1a). The electronic energy levels of $M$ and $X^-$ (referenced with $M^+$ and $X$ for an interionic distance, $R \rightarrow \infty$) against R are schematized in Figure 1a.[1] This is basically an extension of a classical model from refs [2,3]. Isolated ($R \rightarrow \infty$) $M$ and $X$ atoms is the most stable configuration. When the atoms are brought closer, the system prefers the ionic state consisting of $M^+$ and $X^-$ ions, where the $M$ and $X$ levels are inverted because of the Madlung potential. As shown in Figure 1a, the surface ions possess reduced Madlung energy and hence they are separated from their bulk counterparts. Each surface cation or anion forms one surface trap, as suggested by numerical studies.[1] Defining an effective ionic charge ($\delta$), the separated energy levels lie below the CB for $M^{+\delta}$ or above VB for $X^{-\delta}$ which are electron trap ($e$-trap, $E_{sM}$) and hole trap ($h$-trap, $E_{sX}$), respectively (Figure 1a). The depth of the surface-state ($E_{bM} - E_{sM}$ or $E_{sX} - E_{bX}$) is determined by the $|\delta|$ of the $M^{+\delta}$ and $X^{-\delta}$. Contextually, the depth of the trap is deepest for HgS, CdS,[4] and ZnS.[1]

Moving onto the specifics of CdSe quantum dots (QDs), it is an intrinsic $n$-type cubic zinc blend structured crystal under equilibrium, which can take wurtzite structure as well (Figure 1b).[5] VB is formed from 'anions' which are p-like atomic orbital of $Se^{-2}$ ($X^{-2}$), while the bottom of the CB is formed from 'cations', specifically $Cd^{+2}$ ($M^{+2}$).[6] Since the surface adsorbents and passivati-on are ignored the surface states take an ionic state of 'δ' which is the effective ionic charge of the Cd and Se. i.e. in terms of Figure 1b, the surface states are formed from $Cd^{+\delta}$ and $Se^{-\delta}$ ions close to the CB and VB respectively. If more than one δs are present, accordingly a series of surface states can be expected, similar to the case of cation rich material (Figure 1b). Conversely if we consider anion rich material then the surface states are close to the VB formed by $Se^{-\delta}$. Thus the energy gap between the surface states ($E_{sg(\delta)}$) varies depending on the δ. With the presumption that the CB and VB are not broadened, the surface states form discrete levels rather than as continuum of states. Also note that Brus [7] has mentioned about *h*-trapping surface state in QDs.

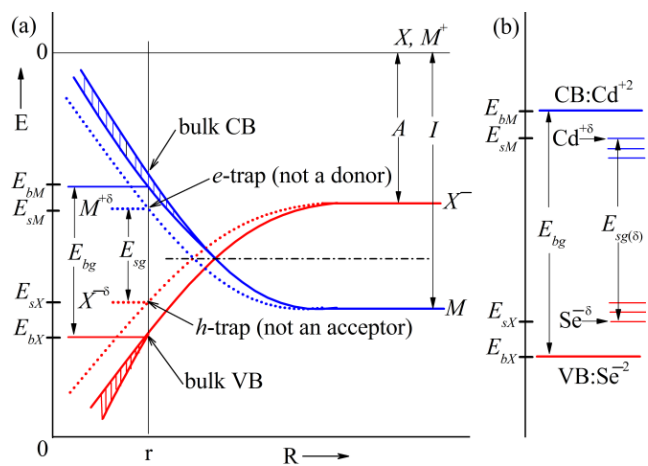

**Figure 1.** Schematic of idealized energy levels (a) in an ionic crystal $M_nX_m$ with reference to interionic distance (R), when R = r it is an equilibrium distance. Band formation is shown in shaded areas, and (b) of CdSe QDs with surface states annotated. Please refer to the text and Appendix A for the meanings of the symbols and associated equations. Part (a) is redrawn based on ref [1].

## II. CONCLUSIONS

This comment not only addresses the fundamental issue but also underscores the importance of the surface states. In CdSe QDs the *h*-traps and *e*-traps are close to the VB and CB, respectively. In anion-rich QD the density of these *h*-trapping surface states increases forming discrete energy levels. If the QD is cation-rich the density of *e*-trapping surface states increases forming discrete energy levels close to the CB. This applies to an ideal scenario of perfect stoichiometry, where the dangling orbitals (depending on the crystal facet) of cation or anion form the surface states or *e*, *h*-traps.

## APPENDIX A

By employing the method of Seitz [2] after Levine et al [1] the following equations can be defined in terms of Madlung potential, electron affinity ($A$) and ionization potential ($I$) where the last two parameters are independent of bulk or surface properties. The mean Madlung potential ($V_b$) of bulk anion ($V_{bX}$) and cation ($V_{bM}$) is given by, $V_b = \frac{1}{2}(V_{bX} + V_{bM})$ which implies the band gap to be $E_{bg} = 2V_b - (I - A)$. Also $E_{bM} = V_{bM} - I$ and $E_{bX} = -V_{bX} - A$. Analogously, the mean Madlung potential ($V_s$) of surface- anion ($V_{sX}$) and -cation ($V_{sM}$) is defined as $V_s = \frac{1}{2}(V_{sX} + V_{sM})$, which implies the energy gap between the surface bands to be $E_{sg} = 2V_s - (I - A)$, in addition to $E_{sM} = V_{sM} - I$ and $E_{sX} = -V_{sX} - A$.

## AUTHOR INFORMATION

**Corresponding Author**

SV: svempati01@qub.ac.uk; fax:(+90) 312 266 4365; tel: (+90) 312 290 3533

**Notes** Author declare no competing financial interest.

## ACKNOWLEDGMENTS

S. V. thanks The Scientific & Technological Research Council of Turkey (TUBITAK) (TUBITAK-BIDEB 2221 – Fellowships for Visiting Scientists and Scientists on Sabbatical) for postdoctoral fellowship. S.V. also thanks Dr. Tamer Uyar, Associate Professor, UNAM-Bilkent University, Ankara, Turkey for hosting in his research group.